\documentclass[12pt]{article}
\usepackage{graphicx}
\begin{document}

\def \d {{\rm d}}

\title{Impulsive waves in de~Sitter and anti-de~Sitter space-times generated
by null particles with an arbitrary multipole structure}

\author{J. Podolsk\'y\thanks{E--mail: {\tt Podolsky@mbox.troja.mff.cuni.cz}}
\\
\\ Department of Theoretical Physics,\\
Faculty of Mathematics and Physics, Charles University,\\
V Hole\v{s}ovi\v{c}k\'ach 2, 18000 Prague 8, Czech Republic.\\
\\
and J. B. Griffiths\thanks{E--mail: {\tt J.B.Griffiths@Lboro.ac.uk}} \\ \\
Department of Mathematical Sciences \\
Loughborough University \\
Loughborough, Leics. LE11 3TU, U.K. \\ }

\maketitle

\begin{abstract}
We describe a class of impulsive gravitational waves which propagate either
in a de~Sitter or an anti-de~Sitter background. They are conformal to
impulsive waves of Kundt's class. In a background with positive cosmological
constant they are spherical (but non-expanding) waves generated by pairs of
particles with arbitrary multipole structure propagating in opposite
directions. When the cosmological constant is negative, they are
hyperboloidal waves generated by a null particle of the same type. In
this case, they are included in the impulsive limit of a class of solutions
described by Siklos that are conformal to {\sl pp}-waves. 
\end{abstract}

\vfil\noindent
PACS class 04.20.Jb, 04.30.Nk

\bigskip\noindent 
Running title: {\it Impulsive waves in (anti-)de~Sitter space-time}

\vfil
\eject

\section {Introduction}
\medskip\noindent
We consider a particular class of exact solutions of Einstein's equations
which describe impulsive gravitational or matter waves in a de~Sitter or
an anti-de~Sitter background. One class of such solutions has recently been
derived by Hotta and Tanaka \cite{HotTan93} and analysed in more detail
elsewhere \cite{PodGri97}. This was initially obtained by boosting the source
of the Schwarzschild--(anti-)de~Sitter solution in the limit in which its
speed approaches that of light while its mass is reduced to zero in an
appropriate way. In a de~Sitter background, the resulting solution describes
a spherical impulsive gravitational wave generated by two null particles
propagating in opposite directions. In an anti-de~Sitter background which
contains closed timelike lines, the impulsive wave is located on a
hyperboloidal surface at any time and the source is a single null particle
with propagates from one side of the universe to the other and then returns
in an endless cycle.

In this paper we investigate a more general class of such solutions. The
global structure of the space-times and the shape of the impulsive wave
surfaces are exactly as summarised above and described in detail in
\cite{PodGri97}. Here we consider a wider range of possible sources. We
present an interesting class of impulsive gravitational waves that are also
generated by null particles, but these particles in general can have an
arbitrary multipole structure. The space-times are conformal to the impulsive
limit of a family of type~N solutions of Kundt's class~\cite{KSMH80}. When
the cosmological constant is negative, the solutions given here can be
related to the impulsive limit of a class of solutions previously given by
Siklos~\cite{Siklos85}.

It may be noted that a family of impulsive spherical gravitational waves have
also been obtained by Hogan \cite{Hogan92}. These are particular (impulsive)
cases of the Robinson--Trautman family of solutions with a cosmological
constant. They will be discussed further elsewhere and are not related to the
solutions given here.

As is well known, the de~Sitter and anti-de~Sitter space-times can naturally
be represented as four-dimensional hyperboloids embedded in five-dimensional
Minkowski spaces. Impulsive waves can easily be introduced into these
space-times using this formalism. This is done is section~2 in which the form
of the solution is constructed explicitly and the nature of its source is
described. Appropriate coordinate systems for the separate cases of de~Sitter
and anti-de~Sitter backgrounds are described respectively in sections 3
and~4 together with a discussion of the geometrical properties of the waves.
Their relation to previously known solutions is indicated in section~5.

\section {An impulsive gravitational wave in a space-time with a
cosmological constant}

We wish to consider impulsive waves in a de~Sitter or an anti-de~Sitter
background. In these cases, the background can be represented as a
four-dimensional hyperboloid 
 \begin{equation}
 {Z_0}^2 -{Z_1}^2 -{Z_2}^2 -{Z_3}^2 -\epsilon{Z_4}^2 =-\epsilon a^2
\label{E2.1} 
 \end{equation}
 embedded in a five-dimensional Minkowski space-time
 $$ \d s^2= \d{Z_0}^2 -\d{Z_1}^2 -\d{Z_2}^2 -\d{Z_3}^2 -\epsilon\d{Z_4}^2 $$ 
 where $a^2=3/\epsilon\Lambda$ for a cosmological constant $\Lambda$,
$\epsilon=1$ for a de~Sitter background ($\Lambda>0$), and $\epsilon=-1$ for
an anti-de~Sitter background ($\Lambda<0$) in which there are two timelike
coordinates $Z_0$ and $Z_4$. Let us now consider a plane impulsive wave in
this 5-dimensional Minkowski background. Without loss of generality, we may
consider this to be located on the null hypersurface given by
 \begin{equation}
 Z_0+Z_1=0, \qquad {Z_2}^2 +{Z_3}^2 +\epsilon{Z_4}^2 =\epsilon a^2. 
\label{E2.5} 
\end{equation} 
so that the surface has constant curvature. For $\epsilon=1$, the impulsive
wave is a 2-sphere in the 5-dimensional Minkowski space at any time $Z_0$.
Alternatively, for $\epsilon=-1$, it is a 2-dimensional hyperboloid. The
geometry of these surfaces has been described in detail elsewhere
\cite{PodGri97} using various natural coordinate systems.

In this five-dimensional notation, we consider the class of complete
space-times that contain an impulsive wave on this background and that can be
represented in the form 
 \begin{eqnarray}
\d s^2&=& \d{Z_0}^2 -\d{Z_1}^2 -\d{Z_2}^2 -\d{Z_3}^2 
-\epsilon\d{Z_4}^2 \nonumber \\
 &&\qquad -H(Z_2,Z_3,Z_4)\delta(Z_0+Z_1)(\d Z_0+\d Z_1)^2   \label{E2.6} 
 \end{eqnarray}
 where $H(Z_2,Z_3,Z_4)$ is determined on the wave surface (\ref{E2.5}). Thus,
$H$ must be a function of two parameters which span the surface. An appropriate
parameterisation of this surface is given by 
 \begin{equation}
 Z_2=a\sqrt{\epsilon(1-z^2)}\cos\phi, \qquad 
Z_3=a\sqrt{\epsilon(1-z^2)}\sin\phi, \qquad Z_4=az \label{E2.7} 
 \end{equation} 
 where $|z|\le1$ when $\epsilon=1$ and $|z|\ge1$ when $\epsilon=-1$. In terms
of these parameters, it can be shown that the function $H(z,\phi)$ must
satisfy the linear partial differential equation 
 \begin{equation}
 (1-z^2)H_{zz}-2zH_z+{1\over1-z^2}H_{\phi\phi} +2H =-\epsilon 8\pi J(z,\phi) 
\label{E2.8}
 \end{equation} 
 where $J(z,\phi)$ represents the source of the wave. It is a remarkable fact
that this equation arises in such a similar form for both de~Sitter and
anti-de~Sitter backgrounds. This equation will be derived separately for both
cases in the following sections.

It may immediately be observed that a solution of (\ref{E2.8}) of the form
$H=$~const. represents a uniform distribution of null matter over the
impulsive surface. This may always be added to any other non-trivial solution.
However, from now on we will only consider solutions which are vacuum
everywhere except for some possible isolated sources.

Let us now consider solutions that can be separated in the form 
 $$ H_m(z,\phi)=H(z)e^{im\phi} $$ 
 where $m$ is a real constant. Since $\phi$ is a periodic coordinate it
follows that, for continuous solutions (except possibly at the poles
$z=\pm1$), $m$ must be a non-negative integer. For a vacuum solution with
this condition, (\ref{E2.8}) reduces to an associated Legendre equation 
 \begin{equation}
 (1-z^2)H_{zz}-2zH_z-{m^2\over1-z^2}H +2H =0. 
 \label{E2.9} 
 \end{equation} 
 This has the general solution 
 $$ H(z)= a_mP^m_1(z) +b_mQ^m_1(z) $$ 
 where $P^m_1(z)$ and $Q^m_1(z)$ are associated Legendre functions of the
first and second kind of degree~1, and $a_m$ and $b_m$ are arbitrary
constants.

The only possible nonsingular solutions involve the associated Legendre
functions of the first kind. These are nonzero here only for $m=0,1$, and the
solutions are given by 
 $$ H(z,\phi)=a_0P^0_1(z)=a_0z, \qquad {\rm and} \qquad
H(z,\phi)=a_1P^1_1(z)e^{i\phi} =-\epsilon a_1\sqrt{|1-z^2|}e^{i\phi} $$ 
 or any linear combination of them. It may then be observed that the second
of the above expressions can be obtained from the first by a simple
``rotation'' of the coordinates on the wave surface (\ref{E2.5}), so that
they are essentially the same solution. We can thus restrict attention to the
space-time (\ref{E2.6}) with $H={a_0\over a}Z_4$. It can then be shown that
this case is conformally flat. The impulsive component in (\ref{E2.6}) can
be removed by the discontinuous linear transformation 
 \begin{eqnarray}
 Z_4&\to& Z_4 -\epsilon{a_0\over2a}U\Theta(U) \nonumber \\
V&\to& V +\epsilon{a_0^2\over4a^2}U\Theta(U)
-{a_0\over a}Z_4\Theta(U) \nonumber
 \end{eqnarray}
 where $U\equiv Z_0+Z_1$, $V\equiv Z_0-Z_1$ and $\Theta$ is the Heaviside
step function. (This does not introduce impulsive components into the Weyl
tensor.) Thus, these nonsingular solutions represent only the de~Sitter or
anti-de~Sitter backgrounds in different coordinates. In these backgrounds,
there is no equivalent to the plane impulsive gravitational wave (for
$\Lambda=0$) for which the Weyl tensor has constant components over the wave
surface.

It now follows that the only nontrivial solution of (\ref{E2.9}) involves the
Legendre functions of the second kind. These necessarily have singularities
at $z=\pm1$ which may correspond to poles at which the sources of the
impulsive wave may be located. Summing over all possible modes, a general real
solution is obtained in the form 
 \begin{equation}
 H(z,\phi)= \sum_{m=0}^\infty b_mH_m(z,\phi)
= \sum_{m=0}^\infty b_mQ^m_1(z)\cos[m(\phi-\phi_m)] \label{E2.10} 
\end{equation}
 where $b_m$ and $\phi_m$ are real constants representing the arbitrary
amplitude and phase of each component. It may be recalled that the associated
Legendre functions of the second kind are generated by the relation 
 $$ Q^m_1(z)=(-\epsilon)^m|1-z^2|^{m/2}{\d^m\over\d z^m}Q_1(z) $$ 
 where $Q_1(z)=Q^0_1(z)$. The first few of these functions are given by 
 \begin{eqnarray}
 Q_1^0(z)&=& {z\over2}\log\left|{1+z\over1-z}\right|-1, \nonumber \\ 
Q_1^1(z)&=& -\epsilon{1\over2}\sqrt{|1-z^2|}\log\left|{1+z\over1-z}\right|
-{z\over\sqrt{|1-z^2|}}, \\ 
Q_1^2(z)&=&\epsilon{2\over1-z^2}, \qquad\qquad
Q_1^3(z)=-{8z\over|1-z^2|^{3/2}} \nonumber \label{E2.11} 
 \end{eqnarray}
 which have been expressed in forms that are applicable for real $z$ for both
$-1\le z\le1$ and $|z|>1$.

The first of these terms ($m=0$) gives the simplest (axially symmetric)
solution
 $$ 
 b_0H_0 ={b_0\over2a}\left[ Z_4\log\left|{a+Z_4\over a-Z_4}\right|-2a
\right]. 
 $$ 
 In fact this is exactly the solution found by Hotta and Tanaka
\cite{HotTan93} (with $b_0=8p$) which was obtained by boosting the source of
the Schwarzschild--(anti-)de~Sitter space-time to the ultrarelativistic
limit. In this case, the singularities correspond to sources represented by
two delta functions 
 $$ 
 b_0 J_0(z,\phi)=\epsilon{b_0\over8\pi} \big[\delta(z-1)+\delta(z+1)\big].
$$ 

Let us now consider the further terms for arbitrary $m$. From the definition
of $H_m(z,\phi)$ given in (\ref{E2.10}) and the identity (\ref{EA.7}) from the
appendix, it can be shown that
 \begin{eqnarray}
 &&(1-z^2)H_{m,zz}-2zH_{m,z}+{1\over1-z^2}H_{m,\phi\phi} +2H_m \nonumber \\ 
&&\qquad=-(-1)^m(1-z^2)^{m/2}
\left[\delta^{(m)}(z-1)+\delta^{(m)}(-z-1)\right]\cos[m(\phi-\phi_m)]  
 \nonumber 
 \end{eqnarray}
 where $\delta^{(m)}$ is the $m^{\rm th}$ derivative of the delta function. 
Comparing this with (\ref{E2.8}), it can be seen that each of the components
$H_m$ corresponds to sources at $z=\pm1$ given by 
 \begin{equation}
 J_m(z,\phi)=\epsilon {(-1)^m\over8\pi} (1-z^2)^{m/2}
\left[\delta^{(m)}(z-1)+\delta^{(m)}(-z-1)\right]\cos[m(\phi-\phi_m)]. 
\label{E2.18} 
 \end{equation} 
 These components describe point sources with an $m$-pole structure. They
have the appropriate dependence on $z$ as the $m^{\rm th}$ derivative of the
delta function, together with the appropriate periodic dependence on $\phi$.
The multipole character of first three of these modes is clearly illustrated
in Fig.~1. (It may be noted that similar multipole sources can generate
impulsive {\sl pp}-waves in space-times with $\Lambda=0$ \cite{GriPod97}.)

\begin{figure}
\begin{center}\includegraphics[scale=0.65, trim=90 80 80 80]{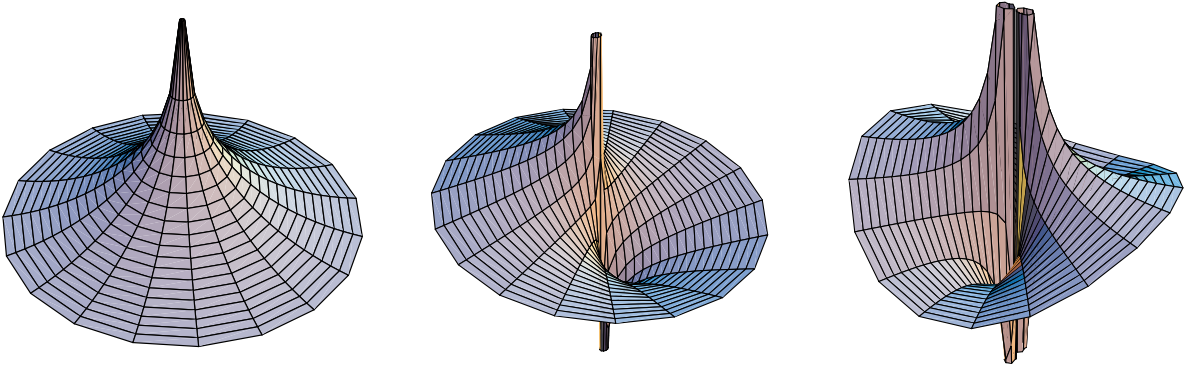}
\caption{ These pictures illustrate the monopole, dipole and
quadrupole modes showing the dependence of the functions $H_0$, $H_1$ and
$H_2$ near the singular point representing one of the sources of the
impulsive waves.} 
 \label{fig:poles}
\end{center}
\end{figure}

Finally we observe that the solution (\ref{E2.10}) represents a general
solution containing point sources which are arbitrary combinations of
$m$-poles 
 $$ 
 J(z,\phi) =\sum_{m=0}^\infty b_mJ_m(z,\phi). 
 $$ 
 The constants $b_m$ represent the strength of each $m$-pole and $\phi_m$ its
orientation.

\section {Impulsive waves in a de~Sitter background} 
\medskip\noindent
When the cosmological constant is positive, it is most convenient to work
with the global coordinate system given by 
  \begin{eqnarray}
 Z_0&=&-a\cot\eta \nonumber \\
Z_1&=&a\,{\rm cosec}\,\eta\cos\chi \nonumber \\
Z_2&=&a\,{\rm cosec}\,\eta\sin\chi\sin\theta\cos\phi 
 \label{E3.1} \\
Z_3&=&a\,{\rm cosec}\,\eta\sin\chi\sin\theta\sin\phi \nonumber  \\
Z_4&=&a\,{\rm cosec}\,\eta\sin\chi\cos\theta \nonumber
 \end{eqnarray}
 in which $\eta,\chi,\theta\in[0,\pi]$, $\phi\in[0,2\pi]$. In these
coordinates it can be seen that the impulsive wave is localised on the
surface given by $\delta(Z_0+Z_1)={1\over a}\delta(\chi-\eta)$. Thus, on the
impulsive null hypersurface $\chi=\eta$ the coordinates (\ref{E3.1}) are
identical to those of (\ref{E2.7}) with the identity $z=\cos\theta$.

In this case the line element (\ref{E2.6}) with the solution (\ref{E2.10})
takes the form 
 \begin{equation}
 \d s^2={a^2\over\sin^2\eta}\bigg(\d\eta^2 -\d\chi^2
-\sin^2\chi(\d\theta^2+\sin^2\theta\d\phi^2)\bigg)
 -aH(\theta,\phi)\, \delta(\eta-\chi)(\d\eta-\d\chi)^2. \label{E3.2}
\end{equation}

Now, in order to justify the equation (\ref{E2.8}), we may adapt the approach
of Dray and 't~Hooft \cite{DrayTH85}. In deriving an exact solution for a
spherical impulsive wave in a Schwarzschild space-time, they have given field
equations for such a wave in a more general class of backgrounds. These also
apply in a space-time with a positive cosmological constant. Using a line
element of the form 
 \begin{equation}
 \d s^2=-2A(u,v)\d u\big[\d v+H(\theta,\phi)\delta(u)\d u\big] 
-g(u,v)(\d\theta^2+\sin^2\theta\d\phi^2) \label{E3.3} 
 \end{equation} 
 and requiring that $A_v=0=A_{vv}$ and $g_v=0=g_{vv}$ on the null
hypersurface $u=0$, the field equations given in \cite{DrayTH85} and
\cite{HotTan93} reduce to the single equation on the impulse 
 \begin{equation}
 \left({A\over g}\Delta-{g_{uv}\over g}\right)\,H =8\pi J(\theta,\phi)
\label{E3.4} 
 \end{equation} 
 where $\Delta$ is the laplacian on the sphere and the source of the wave is
given by $T_{uu}=J(\theta,\phi)\delta(u)$. Now, putting $u\equiv\chi-\eta$
and $v\equiv\cot[{1\over2}(\chi+\eta)]$, the line element (\ref{E3.2}) can be
transformed to the form (\ref{E3.3}) where $A$ and $g$ are given by 
 $$ A(u,v)=-{a^2\over(1+v^2)}{1\over\sin^2(\cot^{-1}v-u/2)}, \qquad
g(u,v)=a^2{\sin^2(\cot^{-1}v+u/2)\over\sin^2(\cot^{-1}v-u/2)}. $$ 
 With these, (\ref{E3.4}) takes the explicit form 
 \begin{equation}
 \left(\Delta+2\right)\,H(\theta,\phi) =-8\pi J(\theta,\phi). 
 \label{E3.5}
 \end{equation} 
 Making the substitution $z\equiv\cos\theta$, the laplacian on a sphere
becomes 
 $$ \Delta =\partial_z\big[(1-z^2)\partial_z\big]
+(1-z^2)^{-1}\partial_\phi\partial_\phi $$
 with which (\ref{E3.5}) takes the form (\ref{E2.8}) which is thus
established for this case.

We now consider the explicit solutions (\ref{E2.10}) given by 
 \begin{equation}
 H(\theta,\phi)= \sum_{m=0}^\infty b_mH_m(\theta,\phi)
= \sum_{m=0}^\infty b_mQ^m_1(\cos\theta)\cos[m(\phi-\phi_m)]. \label{E3.6} 
 \end{equation}

The simplest case in which $b_m=0$ for $m>0$, is exactly the solution 
 $$ 
 H_0={1\over2}\cos\theta
\log\left({1+\cos\theta\over1-\cos\theta}\right)-1 
 $$ 
 obtained by Hotta and Tanaka \cite{HotTan93} as described elsewhere
\cite{PodGri97}. It represents a spherical impulsive wave in a de~Sitter
background generated by two null particles moving in opposite directions. The
particles are situated at the poles $\theta=0$ and $\theta=\pi$. Using these
global coordinates, it can be seen that the impulsive wave is located on the
cosmological horizon of a de~Sitter space-time. (This is analogous to the
solution given in
\cite{DrayTH85} in which the impulsive wave is located on the horizon of a
Schwarzschild space-time.) Moreover, since $\chi=\eta$ on the wave, it can be
seen from (\ref{E3.2}) that at any time the area of the spherical wavefront
spanned by
$\theta$ and $\phi$ is a constant equal to $4\pi a^2$. In fact it describes a
spherical impulsive wave propagating from the North pole to the South pole in
a closed form of the de~Sitter universe which contracts to a minimum size and
then re-expands as described in \cite{HotTan93} and~\cite{PodGri97}.

The general solution (\ref{E3.6}) of the space-time (\ref{E3.2}) can be seen
to represent a similar wave generated by two null particles with arbitrary
multipole structure. The first few higher multipole terms are given simply by
 \begin{eqnarray}
 H_1&=& -\left[
{1\over2}\sin\theta\,\log\left({1+\cos\theta\over1-\cos\theta}\right)
+\cot\theta \right] \cos(\phi-\phi_1) \nonumber \\ 
H_2&=&{2\over\sin^2\theta}\cos[2(\phi-\phi_2)] \nonumber \\
H_3&=&-8{\cos\theta\over\sin^3\theta}\cos[3(\phi-\phi_3)]. \nonumber 
 \end{eqnarray}

It has been argued above that the area of the spherical wavefront spanned by
$\theta$ and $\phi$ is a constant. Therefore this particular wave is
non-expanding (with the background either expanding or contracting through
it). In view of this property, we would expect that this solution can be
related to a particular (impulsive) case of the generalised class of Kundt
waves with non-vanishing cosmological constant $K(\Lambda)$ presented by
Garc\'{\i}a D\'{\i}az and Pleba\'nski \cite{GarPle81}. This has also been
described by Ozsv\'ath, Robinson and R\'ozga \cite{OzRoRo85} as their class
$R(\Lambda,0,1)$. Adapting the coordinate system of \cite{OzRoRo85}, the line
element for this class of solutions can be given in the form 
 \begin{equation}
 \d s^2 =2\left({\xi+\bar\xi\over1+c\xi\bar\xi}\right)^2
\d\tilde u\d\tilde v
-{2\d\xi\d\bar\xi\over(1+c\xi\bar\xi)^2} 
-\left[ 2\left({\xi+\bar\xi\over1+c\xi\bar\xi}\right)^2\tilde v^2
-{\xi+\bar\xi\over1+c\xi\bar\xi}G \right]\d\tilde u^2 
 \label{E22}
 \end{equation} 
 where $c=\Lambda/6$ and $G(\xi,\bar\xi,\tilde u)$ is required to satisfy the
equation 
 \begin{equation}
 {1\over c}(1+c\xi\bar\xi)^2G_{\xi\bar\xi}+2G
={8\pi\over c} {1+c\xi\bar\xi\over\xi+\bar\xi}T_{\tilde u\tilde u}. 
 \label{E23}
 \end{equation} 
 It may be observed that this class is conformal to Kundt's class of type~N
vacuum solutions with vanishing cosmological constant~\cite{KSMH80}. In fact
it can be shown \cite{Podol93} that this is the only class of vacuum
solutions that are conformal to Kundt's class of type~N with $\Lambda$ zero.

Since this is just the de~Sitter space-time when $G=0$ and $\Lambda>0$, we
can concentrate here on the case of an impulsive wave in which
$G(\xi,\bar\xi,\tilde u)=-H(\xi,\bar\xi)\delta(\tilde u)$. Now
performing the transformation 
 $$ 
 \xi=\sqrt2 a \tan{{\theta\over2}} \> e^{i\phi}, \qquad\qquad
 \tilde v={a\over\sqrt2}{1\over t}, \qquad\qquad 
\tilde u={1\over\sqrt2a}(\rho-t), 
 $$ 
 where $a=1/\sqrt{2c}$ in this case, the line element (\ref{E22}) becomes 
 \begin{eqnarray}
 && \d s^2 =a^2t^{-2}\sin^2\theta\cos^2\phi \> (\d t^2-\d\rho^2)
-a^2(\d\theta^2+\sin^2\theta\d\phi^2) \nonumber \\
 &&\qquad\qquad\qquad -\sin\theta\cos\phi \,
H(\theta,\phi)\delta(\rho-t)(\d\rho-\d t)^2. 
 \label{E24}
 \end{eqnarray} 
 This can be seen to be exactly the solution (\ref{E2.6}) in which the
de~Sitter background in the five-dimensional form (\ref{E2.1}) is
parameterised by 
 \begin{eqnarray}
 Z_0 &=& -a\cos\theta +{a^2\over t}\sin\theta\cos\phi
+{\rho^2-t^2\over2t}\sin\theta\cos\phi \nonumber\\
 Z_1 &=& \quad a\cos\theta -{a^2\over t}\sin\theta\cos\phi \nonumber\\
 Z_2 &=& {\rho\over t}\>a\sin\theta\cos\phi \\
 Z_3 &=& \quad a\sin\theta\sin\phi \nonumber\\
 Z_4 &=& \quad a\cos\theta \hskip6pc
-{\rho^2-t^2\over2t}\sin\theta\cos\phi \nonumber
 \label{E25}
 \end{eqnarray} 
 where, for consistency with (\ref{E2.7}) we only need to consider the
impulsive wave located on $\rho=t$. In addition, the field equation
(\ref{E23}) is identical to (\ref{E2.8}).

Finally, we may note that the left hand side of equation (\ref{E3.5}) or
(\ref{E23}) is just the laplacian over the sphere plus two operating on a
function. It follows that the solutions described above can be rotated
arbitrarily over the sphere. Since the equation is linear, solutions can
therefore be constructed which contain an arbitrary number of pairs of
arbitrary multipole particles distributed arbitrarily over the impulsive
spherical wave. However, the impulsive wave is unique --- it is a sphere of
constant surface area equal to $4\pi a^2$.

\section {Impulsive waves in an anti-de~Sitter background} 
\medskip\noindent
When the cosmological constant is negative, it is most convenient to
introduce the global coordinate system given by
 \begin{eqnarray}
 Z_0&=&a(\cosh R+\sinh R\cos\phi)\eta \nonumber \\
Z_1&=&a(\cosh R+\sinh R\cos\phi)\chi \nonumber \\
Z_2&=&a\sinh R\cos\phi 
-{\textstyle{1\over2}}a(\cosh R+\sinh R\cos\phi)(\chi^2-\eta^2) 
\label{E4.1} \\ 
Z_3&=&a\sinh R\sin\phi \nonumber \\
Z_4&=&a\cosh R \hskip2pc
+{\textstyle{1\over2}}a(\cosh R+\sinh R\cos\phi)(\chi^2-\eta^2) \nonumber
 \end{eqnarray}
 in which $\chi,\eta\in(-\infty,\infty)$, $R\in(0,\infty)$ and
$\phi\in[0,2\pi)$. Although this coordinate system is unconventional, it is
particularly convenient for our purposes here. In these coordinates it can be
seen that the impulsive wave is localised on the surface given by
$\delta(Z_0+Z_1)={1\over a}(\cosh R+\sinh R\cos\phi)^{-1}\delta(\chi+\eta)$.
Thus, on the impulsive null hypersurface $\chi+\eta=0$, the coordinates
(\ref{E4.1}) are identical to those of (\ref{E2.7}) with the identity $z=\cosh
R$.

In this case the general line element (\ref{E2.6}) for an impulsive wave in an
anti-de~Sitter background takes the form 
 \begin{eqnarray}
 \d s^2&=&a^2(\cosh R+\sinh R\cos\phi)^2
\big(\d\eta^2-\d\chi^2\big) -a^2\big(\d R^2+\sinh^2R\d\phi^2\big) \nonumber \\
&&\qquad -a(\cosh R+\sinh R\cos\phi)\,H(R,\phi)\,
\delta(\eta+\chi)(\d\eta+\d\chi)^2.  \label{E4.2} 
 \end{eqnarray}
 It may immediately be observed that these coordinates are naturally adapted
such that the impulsive wave is given by $\chi+\eta=0$, and that the wave
surface of a constant negative curvature which is spanned by the parameters
$R$ and $\phi$ do not vary with time. The geometrical properties of these
waves have been described elsewhere \cite{PodGri97} using different
coordinate systems. Basically, the impulsive wave is hyperboloidal and is
generated by a single null particle moving in an anti-de~Sitter background
which contains closed timelike geodesics. The particle propagates from one
side of the universe to the other and then returns in an endless cycle. The
wave propagating in one direction is obtained by the parameterisation
$z=\cosh R$ as above, while propagation in the opposite direction can be
parameterised by changing the signs of $Z_2$, $Z_3$ and $Z_4$ in (\ref{E4.1})
which is equivalent to putting $z=-\cosh(-R)$.

It is also convenient to reparameterise the wave surfaces by introducing an
alternative global coordinate system in which 
 \begin{equation}
 x={a\over\cosh R+\sinh R\cos\phi}, \qquad 
y={a\sinh R\sin\phi\over\cosh R+\sinh R\cos\phi}. \label{E4.3} 
 \end{equation} 
 Then, also putting $\tilde\eta=a\eta$ and $\tilde\chi=a\chi$, (\ref{E4.2})
becomes 
 \begin{equation}
 \d s^2= {a^2\over x^2}\left( \d\tilde\eta^2-\d\tilde\chi^2-\d x^2-\d y^2
-{x\over a}\,H(x,y)\,\delta(\tilde\eta+\tilde\chi)
(\d\tilde\eta+\d\tilde\chi)^2 \right). \label{E4.4} 
 \end{equation} 
 In these coordinates, the parameterisation of (\ref{E2.1}) with
$\epsilon=-1$ is given by 
 \begin{eqnarray}
 Z_0={a\over x}\tilde\eta, \qquad
&&Z_1={a\over x}\tilde\chi,\qquad
Z_2={1\over2x}\big(a^2+\tilde\eta^2-x^2-y^2-\tilde\chi^2\big), \nonumber \\
&&Z_3={a\over x}y, \qquad
Z_4={1\over2x}\big(a^2-\tilde\eta^2+x^2+y^2+\tilde\chi^2\big)
 \end{eqnarray}

It may immediately be observed that (\ref{E4.4}) is conformal to an impulsive
{\sl pp}-wave. In fact it is the impulsive member of a family of solutions
described by Siklos \cite{Siklos85} which include the only vacuum space-times
that are conformal to {\sl pp}-waves. In this work, Siklos found a specific
family of exact type~N solutions (including possible pure radiation) with a
negative cosmological constant given by 
 \begin{equation}
 \d s^2= {a^2\over x^2}\left( \d\tilde\eta^2-\d\tilde\chi^2-\d x^2-\d y^2
-S(x,y,\tilde\eta+\tilde\chi)(\d\tilde\eta+\d\tilde\chi)^2 \right) 
 \end{equation} 
 provided $S$ satisfies the equation 
 \begin{equation}
 S_{xx}+S_{yy}-{2\over x}S_x =-16\pi T_{\tilde u\tilde u}
\label{E4.6} 
 \end{equation} 
 where $\tilde u=\tilde\eta+\tilde\chi$. Since the left hand side does not
depend on $\tilde u$ explicitly, an arbitrary wave profile may be assumed and
the solutions considered here simply correspond to the impulsive case in
which $T_{\tilde u\tilde u} =J(x,y)\delta(\tilde\eta+\tilde\chi)$.
Putting 
 \begin{equation}
 S={x\over a}H(x,y)\delta(\tilde\eta+\tilde\chi), 
 \label{E4.7} 
 \end{equation}
 equation (\ref{E4.6}) can be written as $-x^2(H_{xx}+H_{yy}) +2H =16\pi ax
J(x,y)$ which, using the coordinates $z=\cosh R$ and $\phi$ given by
(\ref{E4.3}), may be confirmed to be exactly of the form (\ref{E2.8}).
Equation (\ref{E2.8}) is thus justified also for the case of a negative
cosmological constant.

Having established the equation (\ref{E2.8}) in this case, we now express
the explicit solutions (\ref{E2.10}) in the form  
 \begin{equation}
 H(R,\phi)= \sum_{m=0}^\infty b_mH_m(R,\phi)
= \sum_{m=0}^\infty b_mQ^m_1(\cosh R)\cos[m(\phi-\phi_m)]. \label{E4.8} 
 \end{equation}
 This now clearly represents an impulsive gravitational wave on a null
hyperboloidal surface generated by a single null particle of arbitrary
multipole structure located at the point $R=0$ on the surface.

As in the previous section, we may finally note that the field equation is
linear and includes the laplacian over a hyperboloidal surface of constant
negative curvature. It therefore again follows that solutions can be
constructed which contain an arbitrary number of arbitrary multipole
particles distributed arbitrarily over the impulsive wave surface.

\section {Further remarks} 
\medskip\noindent
In his paper \cite{Siklos85}, Siklos has also shown that, for the vacuum case
(except for some possible point sources), the general solutions of
(\ref{E4.6}) for $S$ is of the form 
 \begin{equation}
 S=x^2 {\partial\over\partial x} \left({f +\bar f\over x}\right) 
 \label{E5.1}
 \end{equation} 
 where $f=f(\zeta,\tilde u)$ is an arbitrary function of $\zeta=x+iy$ and
$\tilde u=\tilde\eta+\tilde\chi$ (holomorphic in~$\zeta$). For space-times
that are conformal to Kundt waves for both positive and negative cosmological
constant, Ozsv\'ath, Robinson and R\'ozga \cite{OzRoRo85} have presented the
equivalent explicit vacuum solution to their equation (\ref{E23}) which also
involves an arbitrary function which is holomorphic in $\xi$ which is related
to $\zeta$ by 
 $$ \zeta={1-\sqrt{c}\,\xi\over1+\sqrt{c}\,\xi}. $$

Using (\ref{E5.1}) and also (\ref{E4.3}) with $z=\cosh R$, a general vacuum
solution of (\ref{E2.8}) can be written as 
 $$ {1\over a}\, H(\zeta,\bar\zeta) =f_\zeta +\bar f_{\bar\zeta} 
-2{f+\bar f\over\zeta+\bar\zeta} $$ 
 where $f=f(\zeta)$. In terms of the coordinates $z$ and $\phi$, $\zeta$
is given by 
 $$ \zeta =a{1+i\sqrt{z^2-1}\sin\phi\over z+\sqrt{z^2-1}\cos\phi}. $$

The explicit solutions described in the current paper may easily be
represented in this form. For these cases, The Siklos function $S$ may be
expressed as $S=b_m\sum_m S_m(x,y)\delta(\tilde u)$, where $S_m$ corresponds
to the distinct $m$-pole modes $H_m$. For completeness, we may now identify
the expressions corresponding to the first few modes described above.
 \hfil\break For $m=0$, the monopole solution for $H_0$ is
equivalent to 
 $$ 
 S_0(x,y) ={1\over4a^2} \left[ (a^2+x^2+y^2)\log{(x+a)^2+y^2\over(x-a)^2+y^2}
-4ax \right]
 $$ 
 which corresponds to
 $$ 
f_0(\zeta)={1\over4a^2} (\zeta^2-a^2) \log{\zeta+a\over\zeta-a}.  
 $$ 
When $m=1$, the dipole solution for $H_1$ is equivalent to 
 \begin{eqnarray}
 && S_1(x,y)
={1\over4a^2} \left[ \log{(x+a)^2+y^2\over(x-a)^2+y^2}
-{4ax(a^2+x^2+y^2)\over[(x+a)^2+y^2][(x-a)^2+y^2]}\right] \nonumber \\
&&\hskip13pc \times \big[(a^2-x^2-y^2)\cos\phi_1+2ay\sin\phi_1\big] 
 \nonumber
 \end{eqnarray}
 which corresponds to
 $$ 
f_1(\zeta)=-{1\over2a^2} \left[
(\zeta-a)^2e^{i\phi_1}+(\zeta+a)^2e^{-i\phi_1} \right]
\log{\zeta+a\over\zeta-a}.   
 $$ 
 When $m=2$, the quadrupole solution for $H_2$ is equivalent to 
 $$ 
  S_2(x,y) =8ax^3 {\left[(a^2-x^2-y^2)^2-4a^2y^2\right]\cos2\phi_2
+4ay(a^2-x^2-y^2)\sin2\phi_2 \over[(x+a)^2+y^2]^2[(x-a)^2+y^2]^2} 
 $$ 
 which corresponds to
 $$ 
f_2(\zeta)=-a\left[ {1\over\zeta+a}e^{2i\phi_2}+{1\over\zeta-a}e^{-2i\phi_2}
\right]. 
 $$ 

\section*{Acknowledgments}

JP was supported by a visiting fellowship from the Royal Society and, in
part, by the grant GACR-202/96/0206 of the Czech Republic and the grant
GAUK-230/96 of the Charles University.

\section*{Appendix} 

It is well known that, at least in the range $z\in[-1,1]$, any function can
be expressed as a sum of Legendre polynomials. In particular, using the
identity $\int_{-1}^1 P_j(z)Q_1(z)\,\d z={1+(-1)^j\over j(j+1)-2}$, it can be
shown that 
 \begin{equation}
 Q_1(z)=\sum_{j=0}^\infty {j+{1\over2}\over j(j+1)-2}
\left[P_j(z)+P_j(-z)\right]. \label{EA.1} 
 \end{equation} 
 It also follows immediately from the closure property of the set of Legendre
polynomials that 
 \begin{equation}
 \delta(z-1)=\sum_{j=0}^\infty (j+{\textstyle{1\over2}}) P_j(z).
\label{EA.2} 
 \end{equation} 
 The associated Legendre functions are generated by the relations 
 \begin{equation}
 P^m_j(z)\equiv(-1)^m(1-z^2)^{m/2}{\d^m\over\d z^m}P_j(z), \qquad
Q^m_j(z)\equiv(-1)^m(1-z^2)^{m/2}{\d^m\over\d z^m}Q_j(z). \label{EA.3} 
 \end{equation} 
 By differentiating (\ref{EA.1}) $m$ times and multiplying by
$(-1)^m(1-z^2)^{m/2}$, it can be shown that 
 \begin{equation}
 Q_1^m(z)=\sum_{j=0}^\infty {j+{1\over2}\over j(j+1)-2}
\left[P_j^m(z)+(-1)^mP_j^m(-z)\right]. \label{EA.4} 
 \end{equation}

Now, let us introduce the operator $L_m\equiv (1-z^2){\d^2\over\d z^2}
-2{\d\over\d z} -{m^2\over1-z^2}$. Then, applying the identity
$\big[L_m+j(j+1)\big]P_j^m(z)=0$ to (\ref{EA.4}), it can immediately be seen
that 
 $$ 
 \big( L_m+2\big) Q_1^m(z)=-\sum_{j=0}^\infty (j+{\textstyle{1\over2}})
\left[P_j^m(z)+(-1)^mP_j^m(-z)\right]. 
 $$ 
 Using the definition of $P_j^m(z)$ in (\ref{EA.3}), this becomes  
 $$ 
 \big( L_m+2\big) Q_1^m(z)=-(-1)^m(1-z^2)^{m/2}{\d^m\over\d z^m}
\sum_{j=0}^\infty (j+{\textstyle{1\over2}})
\left[P_j(z)+P_j(-z)\right] 
 $$ 
 and, from (\ref{EA.2}), we finally obtain that 
 \begin{equation}
 \big( L_m+2\big) Q_1^m(z)=-(-1)^m(1-z^2)^{m/2}
\left[\delta^{(m)}(z-1)+\delta^{(m)}(-z-1)\right] \label{EA.7} 
 \end{equation} 
 where $\delta^{(m)}$ is the $m^{\rm th}$ derivative of the delta function.

\end{document}